\documentclass{aastex}
\usepackage{spr-astr-addons}
\usepackage{url}

\RequirePackage{color}
\renewcommand{\vec}[1]{\mbox{\boldmath $#1$}}

\newcommand{\emaila}{wilhelm@mps.mpg.de}
\newcommand{\emailb}{bholadwivedi@gmail.com}

\newcommand*\Del{\mathrm{\Delta}}                 
\newcommand{\rmd}{ {\ \mathrm d} }
\newcommand{\uvec}[1]{\hat{\vec #1}}

\begin{document}

\title{Magnetostatics and the electric impact model}
\shorttitle{Magnetostatics and the electric impact model}
\shortauthors{B.N. Dwivedi, H. Wilhelm, and K. Wilhelm}

\author{Bhola N. Dwivedi}
\affil{Dept. of Applied Physics, Indian Institute of Technology
(Banaras Hindu University), Varanasi-221005, India \\ \emailb}
\and
\author{Horst Wilhelm}
\affil{Cephalos Gesell\-schaft f\"ur
Automati\-sie\-rung mbH, Deichstr. 5, 26871 Papenburg, Germany}
\and
\author{Klaus Wilhelm}
\affil{Max-Planck-Institut f\"ur Son\-nen\-sy\-stem\-for\-schung
(MPS), 37191 Katlenburg-Lindau, Germany\\ \emaila}

\today

\vspace{1cm}

\begin{abstract}
The action of certain static magnetic fields on charged test
particles is interpreted as a consequence of the interaction of the
particles with
electric dipole distributions emitted by other charged particles in relative
motion. The dipole model of electric forces
was initially conceived to emulate Coulomb's law, but is applied here to a
wide class of phenomena, such as forces between parallel conductors,
a relativistic correction of Biot--Savart's
law, and the magnetic moment of a current loop.
\end{abstract}

\section{Introduction} 
\label{introd}

An \emph{impact model} for electrostatic forces has been proposed by
\citet{WilDwiWil}
that allows a description of the attraction and repulsion of electrically
charged particles, in line with Coulomb's law,
by the interaction of massless electric dipoles travelling at
the speed of light in vacuum, $c_0$.
We now want to consider some special
cases that would be treated under the heading ``magnetostatic
fields'' in the framework of classical concepts.

\section{Relative motion of charges} 
\label{moving}

In preparation for the following sections, we study a body~B with a
charge~$q$ at rest in an inertial system~${\rm S}'$ (at $x' = 0$).
It moves relative to another body~A with charge $Q$ ($|Q| \gg |q|$) at $x = 0$
of a coordinate system in~S with constant velocity,
$v_{\rm S} = |\vec{v}_{\rm S}|$, parallel to the $x$ axis and an
impact parameter $b > 0$ on the $y$ axis. The discussion is first limited to
the point of closest approach, where the \emph{classical electric field
component} of the charge $Q$ along
$\vec{b} = b\,\uvec{b}$
(unit vector:~$\uvec{b}$) in~S is
%
\begin{equation}
E_b = \frac{Q}{4\,\pi\,\varepsilon_0\,b^2}
\label{restQ}
\end{equation}
with the electric constant in vacuum $\varepsilon_0$.
In system~${\rm S}'$, the Lorentz transformations give
\citep[cf.,][]{Wei34,Gre82,Jac06}:
%
\begin{equation}
E'_b = \gamma\,E_b
\label{restqq}
\end{equation}
as well as
%
\begin{equation}
\vec{B}' =
\frac{\gamma}{c_0^2}\,E_b\,(\vec{v} \times \uvec{b}) =
\gamma\, Q\,\frac{\mu_0}{4\,\pi}
\frac{\vec{v} \times \uvec{b}}{b^2} ~ ,
\label{B_q}
\end{equation}
with $\gamma = (1 - \beta^2)^{-1/2}$ the Lorentz factor
($\beta = v_{\rm S}/c_0 < 1$),
$\vec{B}'$ the magnetic field and $\mu_0$ the permeability of
the vacuum.

Without motion between S and ${\rm S'}$, Coulomb's law could be described
by the dipole model through the exchange of electric dipoles with
momentum $p_{\rm D}$ as:
%
\begin{eqnarray}
p_{\rm D}\,\frac{\Del N_{q,Q}}{\Del t'} =
\frac{|Q|\,|q|}{4\,\pi\,\varepsilon_0\,b^2} =
p_{\rm D}\,\frac{\Del N_{Q,q}}{\Del t} ~ ,
\label{reciprocal}
\end{eqnarray}
with the interaction rates $\Del N_{q,Q}/\Del t'$, $\Del N_{Q,q}/\Del t$
and $t' = t$ in this case.

A relative motion of the bodies~A and B leads to a distinction
between the eigentimes in S and ${\rm S'}$, according to
the special theory of relativity \citep[STR,][]{Ein05}. As seen from $x = 0$,
the eigentime at $x' = 0$ then is according to the
Lorentz transformations
%
\begin{equation}
t' = \gamma\,t
\label{eigentime}
\end{equation}
and the separation distance of~B from~A is
%
\begin{equation}
r' = (b^2 + c_0^2\,\beta^2\,t'^2)^{1/2}~.
\label{distance}
\end{equation}
Integration of the $y$ component of the momentum transfer to~B
along the total path of~$q$ from $t = -\infty$ to $+\infty$ then yields
%
\begin{equation}
(P^{\rm E})_y =
\frac{\gamma\,Q\,q\,b}
{4\,\pi\,\varepsilon_0}\,\int^\infty_{-\infty} \frac{\rmd t}
{(b^2 + c_0^2\,\beta^2\,\gamma^2\,t^2)^{3/2}} ~ .
\label{motiony}
\end{equation}
With the substitution
%
\begin{equation}
X = b^2 + c^2_0\,\beta^2\,\gamma^2\,t^2 ~ ,
\label{substitution}
\end{equation}
the integral is listed in \citet{BroSem} and can be
evaluated as
%
\begin{equation}
\int^\infty_{-\infty} \frac{\rmd t}{X^{3/2}} =
\int^\infty_{-\infty} \frac{\rmd t}{X\,\sqrt{X}} =
\frac{2}{\gamma\,c_0\,\beta\,b^2} ~ .
\label{split}
\end{equation}
Finally, with the help of Eq.~(\ref{restQ}), we get
%
\begin{equation}
(P^{\rm E})_y  =
\frac{Q\,q}{2\,\pi\,\varepsilon_0\,b\,c_0\,\beta} =
\frac{Q\,q}{2\,\pi\,\varepsilon_0\,b\,v_{\rm S}}~ .
\label{transfer}
\end{equation}
The assumption of a
constant $\vec{v}_{\rm S}$ requires special situations
(large mass-to-charge ratios, but see also next section).
In general, the charge~$q$ will be affected by the presence
of $Q$ leading to an elastic
scattering on hyperbolic trajectories (cf., Rutherford's formula).

\section{Parallel conductors} 
\label{conductors}

Another interesting case consists of two (infinitely)
long parallel conductors, ${\rm C}_1$
and ${\rm C}_2$, separated by $b$ and carrying currents,
$I_1$ and $I_2$, respectively.
The classical treatment gives a circular magnetic field around~${\rm C}_1$
produced by the current $I_1$,
which we assume to be much larger than $I_2$.
Applying the law of Biot--Savart, we get
%
\begin{equation}
B_1(b) = \frac{\mu_0}{2\,\pi}\,\frac{I_1}{b}
\label{Ampere}
\end{equation}
at the location of the conductor~${\rm C}_2$, and a force per length $L$ of
%
\begin{equation}
\frac{K}{L} = - \frac{\mu_0}{2\,\pi}\,\frac{I_1\,I_2}{b}
\label{Biot}
\end{equation}
there.

How would the dipole model description cope with this situation?
Let us assume that both conductors and their positive charges, $Q_1$ and $Q_2$,
are at rest in system S, whereas the negative charges in the conduction band
are all in system ${\rm S}'$ and moving with the same speed:
%
\begin{equation}
v_{\rm S} = \frac{1}{Z}\,\frac{\Del L}{\Del t} =
\frac{1}{Z}\,\frac{\Del L'}{\Del t'} =
\frac{1}{Z}\,\frac{\gamma\,\Del L}{\Del t'}> 0 ~ .
\label{speed}
\end{equation}
The number of positive and negative charges, $Z$, per
length, $\Del L$, will be assumed to be the same in both conductors. With
%
\begin{equation}
I_{1,2} = - |q_{1,2}|\,Z\,\frac{v_{\rm S}}{\Del L} ~ ,
\label{current}
\end{equation}
where $-|q_{1,2}|$ are the moving charges,
we get currents in the same direction.
Substitution of $v_{\rm S}$ in Eq.~(\ref{current}) yields
%
\begin{equation}
I_{1,2} = - \frac{|q_{1,2}|}{\Del t} =
- \frac{\gamma\,|q_{1,2}|}{\Del t'} ~ .
\label{current11}
\end{equation}
One can
imagine that the negative charges are fixed
on massless rods. The ``Lorentz contraction''
then leads to a factor of $\gamma$ in the linear number density of the charges
in the moving frame.

We construct,
to demonstrate the principle, our experiment in such a way that the
charges in the conductors are separated by distances much larger
than the separation of the conductors, $b$, i.e.,
we select $b \ll \Del L/Z \ll \Del L$.
In addition, we assume that the pairs of negative or positive charges,
($-|q_1|,-|q_2|$) or ($Q_1,Q_2$),
maintain a constant separation distance of $b$.
Four electrostatic forces have then
to be evaluated between the following pairs:\\
(1) \quad $(Q_1,Q_2)$\\
(2) \quad $(-|q_1|,Q_2)$\\
(3) \quad $(-|q_1|,-|q_2|)$\\
(4) \quad $(Q_1,-|q_2|)$~.\\
It is clear that neither the pairs~(1) and (2) together give a contribution,
nor pairs~(3) and (4), as long as there is no relative
motion. If there is such a motion,
the example in Sect.~\ref{moving} can directly be applied.
According to Eq.~(\ref{transfer}) each isolated pair,
even if its charges are in relative motion,
does not provide a contribution differing from
the electrostatic attraction or repulsion.
Together with Eq.~(\ref{current11}), however,
differential momentum transfers from $- |q_1|$ to $Q_2$
and $Q_1$ to $- |q_2|$
results:
%
\begin{equation}
\Del P =
\frac{q_1\,Q_2}{2\,\pi\,\varepsilon_0\,b\,v_{\rm S}}\,(\gamma - 1) =
\frac{q_2\,Q_1}{2\,\pi\,\varepsilon_0\,b\,v_{\rm S}}\,(\gamma - 1) ~ .
\label{Ppair}
\end{equation}
The evaluation of Eq.~(\ref{Ppair}) will be carried out
for the length $\Del L$ in three steps:
\begin{itemize}
\item The charges
$-|q_1|$ move with $v_{\rm S}$ in case~(2).
We consider (as an approximation under our geometric assumptions)
the interaction with $Z$ charges $Q_2$
to be the sum of Z individual interactions of pairs $(- |q_1|, Q_2)$ per
length, $L$, and get
%
\begin{equation}
\frac{P_2}{\Del L} = Z\,\frac{\Del P_2}{\Del L} =
-\frac{(Z/\Del L)\,|q_1|\,Q_2}
{2\,\pi\,\varepsilon_0\,b\,v_{\rm S}}\,(\gamma - 1)~ .
\label{Npair}
\end{equation}
Since $\gamma$ can be approximated for small $\beta$ by the first two terms of
the expansion
%
\begin{equation}
\gamma = 1 + \frac{1}{2}\,\beta^2 + ... ~ ,
\label{Gamma}
\end{equation}
we find
%
\begin{equation}
\frac{P_2}{\Del L} \approx
-\frac{(Z/\Del L)\,|q_1|\,Q_2\,v_{\rm S}}{4\,\pi\,\varepsilon_0\,b\,c_0^2} ~.
\label{Lpair}
\end{equation}
Considering that the factor $\gamma$ in Eq.~(\ref{current11}) has been taken
care of in Eq.~(\ref{Ppair}) and
that $- (Z/\Del L)\,Q_2\,v_{\rm S}$ can
be expressed by $I_2$ in conductor C$_2$,
half the specific force can be written as
%
\begin{equation}
\frac{K_2}{\Del L} = \frac{P_2}{\Del L\,\Del t} \approx
- \frac{\mu_0\,I_1\,I_2}{4\,\pi\,b} ~ ,
\label{curorhalf}
\end{equation}
noting that $c^2_0 = (\mu_0\,\varepsilon_0)^{-1}$.

\item The symmetry between cases~(2) and (4) in Eq.~(\ref{Ppair})
then yields for parallel currents
%
\begin{equation}
\frac{K_{\rm p}}{\Del L} =
\frac{K_2 + K_4}{\Del L} \approx - \frac{\mu_0\,I_1\,I_2}{2\,\pi\,b} ~ ;
\quad I_1\,I_2 > 0 ~ .
\label{curforce}
\end{equation}

A full treatment without approximations will have to confirm
that Eq.~(\ref{curforce}) is valid as an equality.

\item A slight complication in our argumentation arises
when we consider opposite currents in parallel conductors.
The negative charges moving relative to the positive ones in the other
conductor lead to the same attraction as in the previous example,
namely $(K_2 + K_4)/\Del L$. However, the negative
charges in the conductors are now counter-moving with a relative speed
of $2\,v_{\rm S} \ll c_0$. So, we have for these charges
%
\begin{equation}
\gamma' \approx 1 + \frac{2\,v_{\rm S}^2}{c_0^2}+...
\label{gamma_p}
\end{equation}
and an additional specific force
%
\begin{equation}
\frac{K_3}{\Del L} \approx
- \frac{\mu_0\,I_1\,I_2}{\pi\,b} \quad ; \quad I_1\,I_2 < 0 ~ .
\label{addforce}
\end{equation}
This repulsion is twice as large as the attraction.
Conceptually, this result could have been obtained by assuming that
the negative charges are moving in one conductor and the positive charges
in the other one.

\end{itemize}

Thus the dipole model gives an expression for the specific force
equivalent to that of the classical
magnetostatic field in Eq.~(\ref{Biot}) for both parallel and
antiparallel currents.

\section{Charge near a current} 
\label{charge}

Assume moving negative charges causing a current, $I$, in
a conductor, C, which is at rest in system S,
together with its positive charges,
and a negative test charge $-|q| = -|e|$ at a distance
$b$ from C. Biot--Savart's law
gives a magnetic field according to Eq.~(\ref{Ampere}) at the
position of the test charge. If this is at ``rest'', no Lorentz force would be
expected. The dipole model provides a clear answer when this condition is
fulfilled, namely, when the test charge is moving
parallel to the conductor with $v_{\rm S}/2$, half the speed of the negative
charge carriers in the conductor, because then the effects of its
positive and negative charges on the test charge cancel each other.
This can be seen as a relativistic correction of the law of
Biot--Savart \citep[cf.,][]{Gre82}.

As an example, we consider a long conductor with a cross-section, $A$.
In a certain volume, $V = A\,\Del L$,
with length $\Del L$, there are $Z$ electrons as current carriers available.
They are at rest in system ${\rm S}'$ moving with
%
\begin{equation}
v_{\rm S} = - \frac{\Del L}{Z}\,\frac{I}{|e|}
\label{espeed}
\end{equation}
relative to the positive charges, cf., Eq.~(\ref{current}). If we place the test
charge at $b \ll \Del L$ from the conductor, the momentum transfer will
not be significantly affected by charges outside $\Del L$, and we can
divide Eq.~(\ref{Ppair}) by $\Del t = \Del L/(Z\,v_{\rm S})$
to find the force acting on the
test charge (applying arguments similar to those of the previous section):
%
\begin{equation}
K = \frac{\Del P}{\Del t}
\approx -\frac{\mu_0\,|e|\,I\,v_{\rm S}}{4\,\pi\,b} ~ .
\label{current3}
\end{equation}
Approximately
$8.52 \times 10^{21}$ electrons are in the conduction band
of copper for $A = 0.1~{\rm mm}^2$ and $L = 1$~m \citep{Wes56}.

A current of $I = 1$~A then requires $v_{\rm S} = 0.73~{\rm mm\,s}^{-1}$, and
an electron (at rest in S) at $b = 1$~cm experiences
a repulsive force $K \approx 1.2 \times 10^{-27}$~N.
A smaller cross-section of $A = 0.001~{\rm mm}^2$, for instance,
would then lead to an increase of the quantities $v_{\rm S}$ and $K$
by a factor of one hundred.

The significant feature of Eq.~(\ref{current3})
is that the force $K$ depends on
the speed of the charge carriers, and therefore on the characteristics of the
conductor. This result might be used as an experimental test in a suitable
arrangement.

\section{Magnetic moment} 
\label{Mmoment}

A magnetostatic dipole field can also be treated in this context.
A current, $I$, in a circular conductor with
radius, $R$,
will generate a magnetic moment $\mu = \pi\,R^2\,I$. Biot--Savart's law
gives a magnetostatic field in a point~O
on the dipole axis
in figure~1
of
%
\begin{equation}
\vec{B} =
\frac{\mu_0}{2\,\pi}\,\frac{\mu}{b^3}\,\uvec{z} ~ ,
\label{mag_mom}
\end{equation}
with
$b = |\vec{b}_+| = |\vec{b}_-| = (z^2 + R^2)^{1/2}$
in the near-field region \citep[cf., e.g.,][]{Jac06}.
If a body with charge, $|q|$, is at rest there,
no force will act on it.
However, a Lorentz force would influence a charge moving with the
velocity, $|\vec{v}| > 0$.
Let us assume that the velocity is parallel
to the $y$ axis pointing into the plane of the paper.
%
\begin{figure}[t]
\begin{center}
\includegraphics[width=\columnwidth]{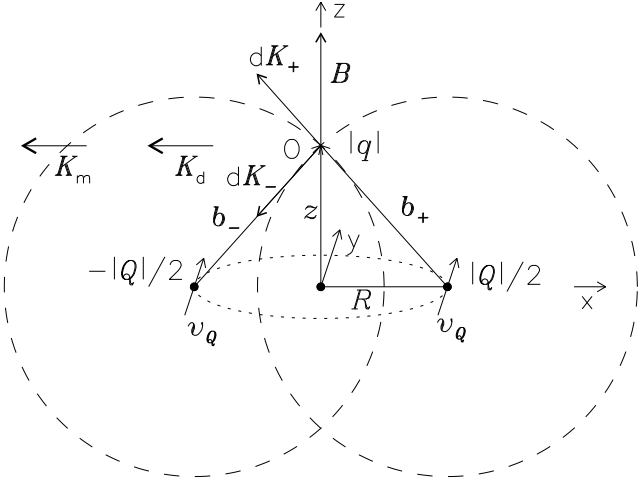}
\end{center}
\caption{Diagram illustrating the magnetic moment discussion.
The angle between $R$ and $\mbox{\boldmath$b_+$}$ is defined as $\theta$, and
the central angle of the dotted circle as $\varphi$.
\label{Moment}}
\end{figure}
A force
%
\begin{equation}
\vec{K}_{\rm m} =
-\frac{\mu_0}{4\,\pi}\,\frac{R\,v^2}{b^3}\,|q|\,Q\,\uvec{x}
~ ,
\label{force_m}
\end{equation}
would result, where we have used a current
$I = Q/T = Q\,v_{\rm Q}/(2\,\pi\,R)$
generated by a charge, $Q$, moving around the dotted circle with
a period, $T$, so that, for reasons of simplicity,
its speed is $v_{\rm Q} = v = |\vec{v}|$.

We have split up the moving charge
into $\pm\,|Q|/2$ as a trick to eliminate the averaged electrostatic force.
We will first discuss two special cases in the framework of
the dipole model\footnote{We again assume the masses of the bodies involved
to be so large that their accelerations can be neglected.}:
(a) Let $|q|$ move with the same velocity vector as
indicated for $\pm\,|Q|/2$. Then there are no forces other than the
electrostatic ones. (b) If $|q|$ moves
antiparallel, we get $\beta' = 2\,v/c_0$ and additional forces.

The equivalence of the
force resulting from the dipole model with Eq.~(\ref{force_m})
can now be shown by
integrating the movements of the charges $\pm\,|Q|/2$ over a full revolution.
With the help of Eqs.~(\ref{restQ}) and (\ref{restqq})
(for which we have shown the equivalence with the dipole model), applied
separately to $\pm\,|Q|/2$,
we find for the force averaged over a period $T$:
%
\begin{eqnarray}
\langle \vec{K}_\pm \rangle =
\frac{|q|}{T}\,\uvec{b}_\pm \int^T_0 \gamma\,E_\pm \rmd t =
\frac{\pm\,|Q|}{8\,\pi\,\varepsilon_0\,b^2}\frac{|q|}
{T}\,\vec{b}_\pm\,\int^T_0 \gamma\,\rmd t ~ .
\label{revolution}
\end{eqnarray}
Substituting in Eq.~(\ref{Gamma})
%
\begin{equation}
\beta^2(\varphi) = \frac{4\,v^2}{c_0^2}\,\cos^2\frac{\varphi}{2} ~ ,
\label{beta}
\end{equation}
obtained from the vector sum of the velocities of $q$ and $Q$,
and replacing $\rmd t$ by the azimuthal angle
%
\begin{equation}
\rmd \varphi = \pm\,v\,\rmd t/R
\label{revol}
\end{equation}
for the positive charge or the negative one, respectively,
we get with $\cos \theta =R/b$ after a straight forward calculation,
taking into account
Eqs.~(\ref{speed}) and (\ref{Gamma}),
the total average force
exerted by the electric dipoles
%
\begin{equation}
\vec{K}_\rmd = (\langle \vec{K}_+ \rangle +
\langle \vec{K}_- \rangle)\,\cos \theta =
-\frac{\mu_0}{4\,\pi}\,\frac{R\,v^2}{b^3}\,|q|\,Q\,\uvec{x}
~ ,
\label{force_d}
\end{equation}
which is the same as $\vec{K}_{\rm m}$ in Eq.~(\ref{force_m}).

It should be noted that this integration involves accelerations of the charges
in the conductor, therefore, we can only expect an estimate for
small accelerations, i.e., for large $R$.
How these considerations have to be applied to magnetic moments of charged
particles is, of course, of great interest, but is beyond the scope of
this conceptual presentation (for a recent discussion, see \citet{HugKin}.

\section{Discussion and conclusions} 
\label{concl}

The electric impact model not only describes electrostatic
forces \citep{WilDwiWil}, but also forces acting on moving charges
seemingly caused by static magnetic fields.
In Sect.~\ref{conductors}, the speed of the negative charge
carriers did not enter into the final result and thus the
definition of the Syst\`eme International d'Unit\'es
\citep[SI;][]{BIPM} of the unit of the electric current is not affected:

The ampere is that constant current which, if maintained in two straight
parallel conductors of infinite length, of negligible circular cross-section,
and placed 1~m apart in vacuum, would produce between these conductors a force
equal to $2 \times 10^{-7}$~N per metre of length.
This definition leads to $\mu_0 = 4\,\pi \times 10^{-7}$~H\,m$^{-1}$ (exact).

On the other hand, it has been found in Sect.~\ref{charge}
that the application of Biot--Savart's
law in determining the Lorentz force acting on a charged particle depends on
the speed of the charge carriers according to STR and, therefore, on the
characteristics of the conductor.


\begin{acknowledgements}
We thank
Eckart Marsch,
Harry Kohl,
Luca Teriaca and
Werner Curdt
for many discussions on these topics. Their critical comments have been
very helpful in formulating our ideas. This research has made extensive use of
the Astrophysics Data System (ADS).
\end{acknowledgements}

\end{document}